# Quantum Gates and Quantum Circuits of Stock Portfolio

## Ovidiu Racorean

*Financial Research department, Ministry of Public Finance, State Treasury, Bucharest, Romania*

**Abstract**

In quantum computation, series of quantum gates have to be arranged in a predefined sequence that led to a quantum circuit in order to solve a particular problem. What if the sequence of quantum gates is known but both the problem to be solved and the outcome of the so defined quantum circuit remain in the shadow? This is the situation of the stock market. The price time series of a portfolio of stocks are organized in braids that effectively simulate quantum gates in the hypothesis of Ising anyons quantum computational model. Following the prescriptions of Ising anyons model, 1-qubit quantum gates are constructed for portfolio composed of four stocks. Adding two additional stocks at the initial portfolio result in 2-qubit quantum gates and circuits. **Hadamard gate, Pauli gates** or **controlled-Z gate** are some of the elementary quantum gates that are identified in the stock market structure. Addition of other pairs of stocks, that eventually represent a market index, like *Dow Jones industrial Average*, it results in a sequence of n-qubit quantum gates that form a quantum code. Deciphering this **mysterious quantum code of the stock market** is an issue for future investigations.

**Key words:** braid of stocks, quantum gates, stock market quantum code, quantum circuits, topological quantum computation.

*e-mail address: decontatorul@hotmail.com

# 1. Introduction

The present paper reveals the interdependencies between different time series by the braids that it forms in time. Applied to time series of stocks prices the braids that result prove to be directly connected to topological models of quantum computation. Eventually the evolution in time of a portfolio of stocks can be represented in a sequence of quantum gates like:

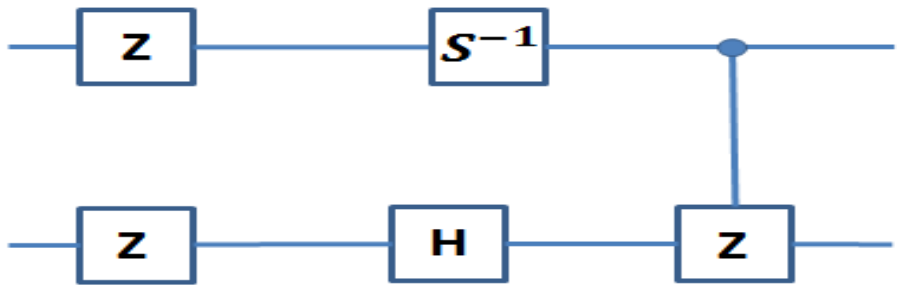

Although it is hard to imagine now that the quantum circuit above is equivalent with the more conventional chart representation of cumulative stocks time series below,

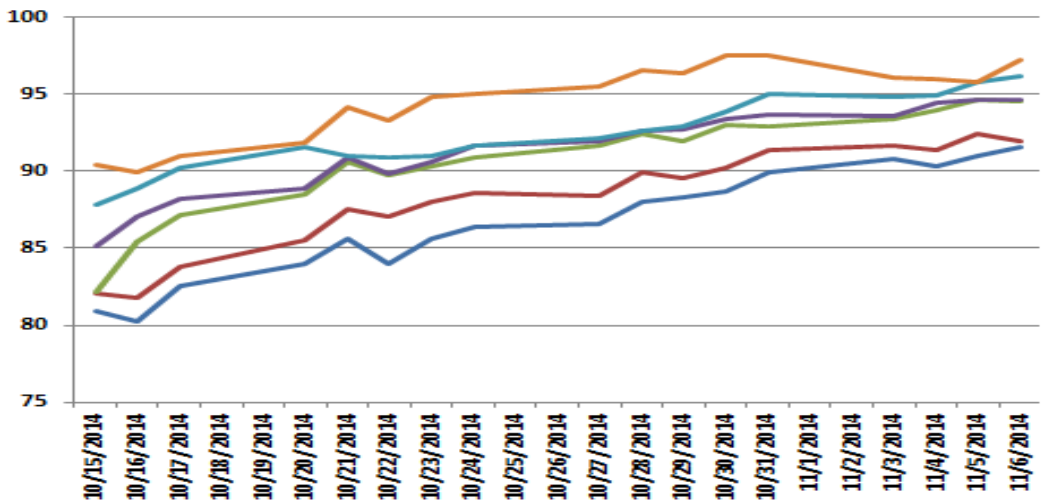

at the end of the presentation this association of images will become natural.

A series of recent papers [15], [16], [17] formally introduced and developed the braid diagrams of stocks prices time series in a portfolio and sketched their connection with the topological methods of quantum computation.

In topological quantum computation implementation of quantum gates is based on the braiding trajectories of some exotic particles, called non-abelian anyons. In the braid generated by trajectories of particles every crossing (under/ over-crossing) of adjacent strands is expressed as a matrix. Joining a predefined sequence of braid matrices together implements a certain quantum gates.

The form of the so constructed elementary braid matrices depends on the class of particles the topological quantum computation is based on. The best suited topological quantum computation model to perform simulations over stock portfolios seems to be the Ising anyons model. Unlike Fibonacci anyons model [2], [11], the quantum computation with Ising model necessitates a smaller number of braid matrices to realize a quantum gate.

To simulate the braid of stocks a quasi-particle must be assigned to each and every stock component of a portfolio. The quasi-particles are manipulated in braids according to stocks price quotations and the sequences of resulting braid matrices are joined in quantum gates.

The stocks subjected to quantum computation simulation are selected from **Dow Jones Industrial Average (DJIA)** components and their price time series are formed by daily closing prices over a period of 1 year in 2014. More spectacular braid structures can appear by using smaller time frames (minutes or even tick-by-tick data).

Construction of quantum gates from the elementary braid matrices of the stocks prices closely follows the work of L. S. Geogiev [8], [9], [10]. Following the line of exposing the problematic of Ising model in his work, first the most basic quantum gates are implemented for single-qubit systems.

The **1-qubit** systems are realized in the Ising anyons model by 4 quasi-particles such as, to perform the simulation, a portfolio of 4 stocks is needed. For a selection of 4 stocks, **AXP, UNH, PG** and **DIS**, a series of elementary quantum gates **like Hadamard gate, S-gate or X-gate** are found to encode the portfolio evolution.

Assembling these quantum gates together it results **a quantum circuit**. Typically quantum circuits are designed to solve certain problems, but the problem to be solved and the output answers of the **stock market quantum circuits** are uncertain.

Complexity of quantum gates increase once the number of quasi-particles extends to 6. The system of 6 Isisng anyons can implement a **2-qubit** basis. A portfolio of 6 stocks must be selected to construct 2-qubit quantum gates. The selected portfolio is composed of **AXP, UNH, DIS, NKE, MCD,** and **HP.** This time, in addition to 1-qubit gates, a **controlled-Z** is realized.

Joining the quantum gates implemented for the 6 stocks portfolio results the quantum circuit in the figure depicted at the starting of this introductory section. The revealed quantum circuit is a quantum code that can be **simulated by any type of quantum computer.**

The scheme of finding quantum gates for stock portfolios suits well to larger number of qubits that could even represent a market index, like *Dow Jones Industrial Average*. The n-qubit quantum gates that can be realized for DJIA form the quantum code of the stock market.

The issue in the case of stock portfolio is that the quantum gates implemented by braiding the price time series are **arbitrary,** are not meant to solve a particular problem. The sequence of quantum gates, in the case of the stock market, defines a *quantum code that remains mysterious* for now.

## 2. Braiding the prices of stocks

The braid representation of prices time series of stocks in a portfolio prove to be a promising candidate in constructing a basis platform quantum computational methods could be applied on. Present paper extends the findings of the previous work on the quantum computation applications in finance [17] by realizing a series of elementary quantum gates for braided stocks of a portfolio.

In illustrating the building of braid diagrams for portfolios, stock components of *Dow Jones Industrial Average (DJIA)* are considered. The selection of stocks that form the portfolio is not restricted at DJIA components; any other stocks might be chosen. The only constrained in selecting the stocks is that their prices must have close values. This limitation in selecting the stock portfolio will be explained later in the section.

Construction of stocks prices braid diagrams standing at the foundation of quantum gates realization is intuitively easy and in what follows it will be exemplified by considering the stocks:

**The Procter & Gamble Company (PG)**, **The Walt Disney Company (DIS)**, **Nike Inc. (NKE)**, **McDonald's Corp. (MCD)**, **American Express Company (AXP)**, **UnitedHealth Group Incorporated (UNH)**, **The Home Depot, Inc. (HD).**

The prices of stocks that will be used further consist of historical quotations of daily closed prices over a period of 1 year in 2014, taken from Yahoo finance. The time frame was selected mainly for the availability of data, more spectacular braid structures can appear by using smaller time frames (minutes or even tick-by-tick data).

A fragment time interval of the selected stocks prices time series, from 3/19/2014 to 3/24/2014 is arranged in the Table 1. The prices time series of stocks are arranged in columns having prices ordered from the smallest (**PG**) on the left to the bigger (**MCD**) on the right at 3/19/2014.

| | | | | | | | |
|---|---|---|---|---|---|---|---|
| 3/24/2014 | 79.3 | 74.86 | 79.66 | 81.28 | 79.49 | 91.01 | 96.18 |
| 3/21/2014 | 77.88 | 75.21 | 80.42 | 81.34 | 80.35 | 91.52 | 95.47 |
| 3/20/2014 | 78.32 | 79.27 | 80.09 | 81.53 | 80.81 | 91.69 | 96.6 |
| 3/19/2014 | 78.78 | 79.15 | 79.75 | 79.96 | 80.52 | 90.73 | 96.1 |
| | pg | nke | hd | unh | dis | axp | mcd |
| | 1 | 2 | 3 | 4 | 5 | 6 | 7 |

**Table 1.** Prices time series of stocks are arranged in columns having prices ordered from the smallest (PG) on the left to the bigger (MCD) on the right

It is important to note here that price time series of stocks are colored; every time series having its own color. This feature is intended to help tracking the trajectories of prices on the next step, where all the rows of the table follows the same rule of prices arrangement in ascending order from the left to the right. The result of the so organized prices time series is shown in the Table 2, below.

| | | | | | | | |
|---|---|---|---|---|---|---|---|
| 3/24/2014 | 74.86 | 79.3 | 79.49 | 79.66 | 81.28 | 91.01 | 96.18 |
| 3/21/2014 | 75.21 | 77.88 | 80.35 | 80.42 | 81.34 | 91.52 | 95.47 |
| 3/20/2014 | 78.32 | 79.27 | 80.09 | 80.81 | 81.53 | 91.69 | 96.6 |
| 3/19/2014 | 78.78 | 79.15 | 79.75 | 79.96 | 80.52 | 90.73 | 96.1 |
| | pg | nke | hd | unh | dis | axp | mcd |
| | 1 | 2 | 3 | 4 | 5 | 6 | 7 |

**Table 2**. The winding trajectories of stocks prices time series that result ordering the prices ascending from the left to the right

It is already easy to notice the winding trajectories of stocks prices following the colored prices time series in the Table 2. A more illustrative picture of winding trajectories is realized by skipping the prices of stocks and focusing only on the prices trajectories as is shown in the figure 1.

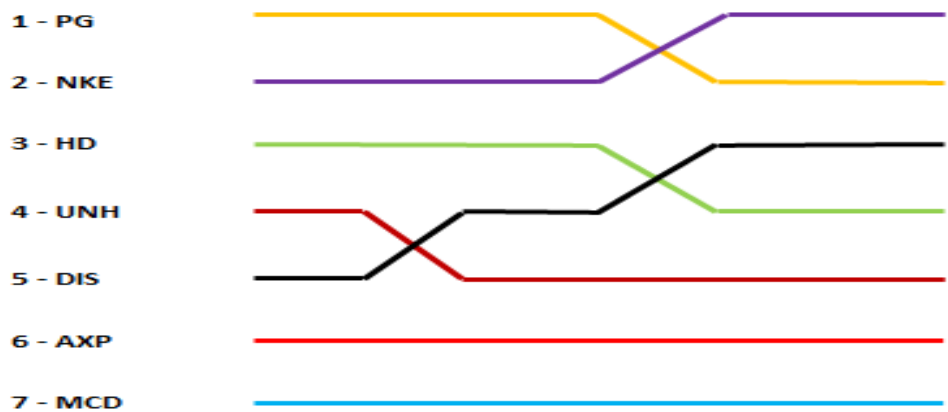

Figure 1. Crossing stocks diagram of stocks prices time series

The resulting diagram is called **crossing stocks diagram** since it clearly points out the intersections of adjacent stocks prices. The diagram of the winding trajectories of stocks prices is ported horizontally for later convenience. Here the time flows from the left to the right.

There is another ingredient that should be added here to complete de picture. Notice that at every crossing of stocks (every intersection of adjacent stocks) the price of a stock came over/under the price of its neighbor. It cannot be said only by analyzing the crossing stocks diagram which one is the stock having the significant bigger moving of price comparing with its neighbor.

To quantify this important information into the **crossing stocks diagram** a simply operation is perform at this point. Every crossing of two adjacent stocks is evaluated as the difference between the price after and before the crossing. The differences are taking in modulo since only the net amounts are considered, such that:

$$\Delta_{Stock\ i} = |P_{before\ crossing} - P_{after\ crossing}|, \qquad (1)$$

$$\Delta_{Stock\ i+1} = |P_{before\ crossing} - P_{after\ crossing}|, \qquad (2)$$

where P is the price of the stock. The stock with the higher difference it will come over and the stock with the smaller difference will be under in a stock crossing.

The two cases that can arise are:

- $\Delta_{Stock\ i} > \Delta_{Stock\ i+1}$ - the stock $i$ is crossing over the stock $i + 1$, in which case the stocks crossing will be called to be an ***overcrossing of stocks***,
- $\Delta_{Stock\ i} < \Delta_{Stock\ i+1}$ - the stock $i$ is crossing under the stock $i + 1$, in which case it will be called to be an ***undercrossing of stocks***.

The two separate cases for the crossing of two stocks are exemplified in the figure 2.

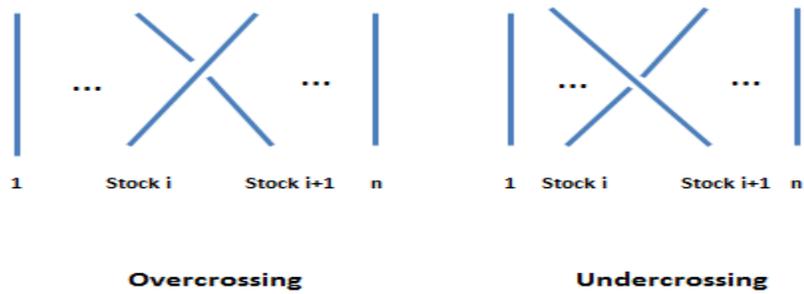

**Figure 2.** Overcrossing and undercrossing of stocks

Calculating the delta for all crossings in the figure 1, it result the braid diagram shown in the figure 3. More in depth details on calculating the deltas between prices of stocks in a crossing can be found on [16] and [17].

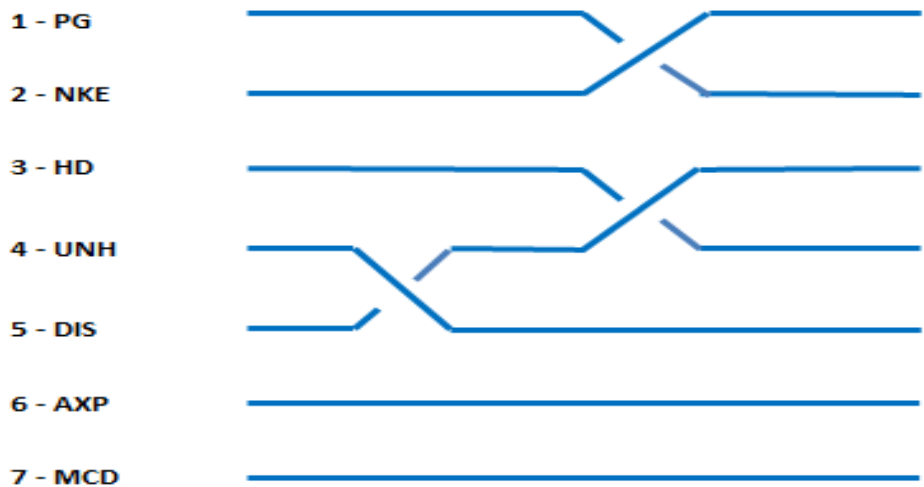

**Figure 3**. Braid diagram of stocks

The colored trajectories of prices are no longer necessary since the only important aspect related to braid diagrams is the over-crossings and under-crossings.

The diagram above is named after its more known counterpart in mathematics. The **braid theory** first introduced in the work of E. Artin [1] has known a flourishing period in the 80s once it was successfully applied to quantum field theory in physics. The late 90s and early 2000s added more shining to braid theory through its surprising applications to quantum computing. The way braid theory is applied to quantum computation will be discussed in the next section; here a quick remainder of some important features of braids should be discussed.

Returning to the braid of stocks, some well-known notations from braid theory should be mentioned here. The over-crossings and under-crossings in the strands of a braid also called **generators of the braid** are noted in figure 4 at the bottom, as $\sigma$ respectively $\sigma^{-1}$. The under numbering that can be observed at the braid generators expressed the number of strand in the braid where an overcrossing and respectively an undercrossing arise. In the figure 4 $\sigma_4$ means that an overcrossing appears on the braid at strand 4 (**UNH**); the braid generator $\sigma_1^{-1}$ signify that an undercrossing arise on the braid at the first strand (**PG**).

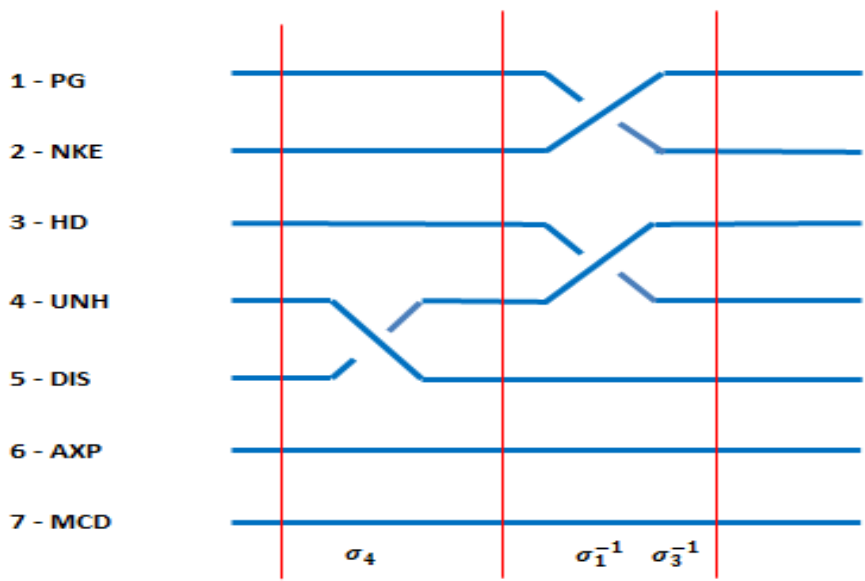

**Figure 4**. The generators of the braid of stocks

Prepared this way the multivariate time series expressing the prices of stocks in a portfolio is ready to be simulated in quantum computation.

## 3. Topological Quantum Computing

The braid diagrams of stock prices constructed in the latter section open the gate to harness the remarkable power of quantum computation to financial applications. The appropriate quantum computational model to analyze the braiding of stock prices time series is the *topological quantum computation model*.

The topological quantum computational model executes universal quantum gates by braiding the trajectories of exotic particles, called anyons, which live in 2-dimensional plane. The specification of planarity for the position of quasi-particles is crucial since by swapping their position in plan, trajectories of quasi-particles executes braids that can be compared further with the prices of stocks braids. Figure 5 depicts the exchange of positions for two quasi-particles in a 2-dimensional plane (on the left) and the braiding trajectories that result (on the right).

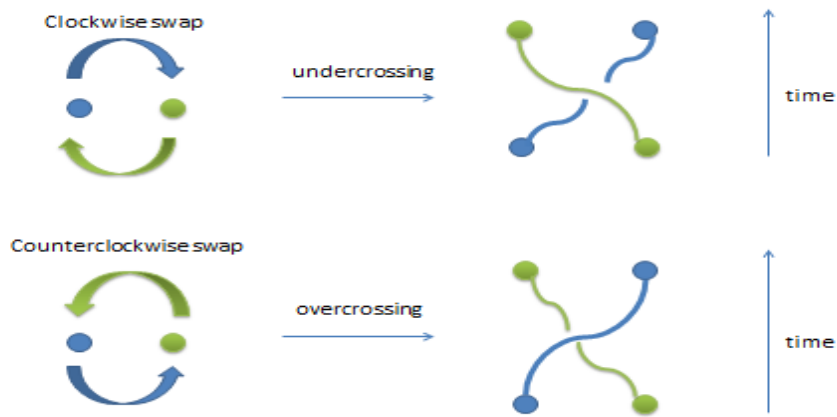

**Figure 5**. Non-abelian anyons exchanging places and their braiding trajectories. A clockwise swap is similar to an undercrossing and counterclockwise swap is reflected in an overcrossing

It can be noticed from figure 5 that a clockwise swap is reflected in an overcrossing and a counterclockwise swap in an undercrossing, of a braid representation. The two possible anyons paths are topologically distinct. To distinguish between these two possible quantum states, the topological quantum computer needs a particular type of anyons, called **non-abelian anyons**. Non-abelian anyons have the property that the order in which particles are exchange positions is important; it said that exhibit braiding statistics.

Not delving in technical details, that can be found in [6], [7], it is suffice to point out here that the final state $\Psi_f$ for a quantum system consisting in a pair of non-abelian anyons $a$ and $b$, having the initial state defined by the wave function $\Psi_i$, differs as follow:

- For a clockwise swap of positions - $\Psi_f \to M_{ab}\Psi_i$;
- For a counterclockwise swap - $\Psi_f \to N_{ab}\Psi_i$.

The matrices $M_{ab}$ and $N_{ab}$ do not commute, such that:

$$M_{ab} \neq N_{ab} . \tag{1}$$

The matrices $M_{ab}$ and $N_{ab}$ are called **elementary exchange matrices** (or **braid matrices**) and are fundamental for the further discussion. The braid matrices are the expression of braid generators for the braiding trajectories of quasi-particles. The braid matrices reflect the overcrossings and undercrossings on the braid resulting from trajectories of quasi-particles that exchange their positions counterclockwise and, respectively clockwise in a plane. Considering the braid generators notation mentioned in the latter section, the elementary exchange matrices can be written as:

$$\rho(\sigma) = N_{ab}$$

$$\rho(\sigma^{-1}) = M_{ab} \tag{2}$$

The above notation of braid matrices also give the hint that matrix $M_{ab}$ actually is the inverse of the matrix $N_{ab}$. This is an important result which leads to the conclusion that all braids of quasi-particles trajectories can be constructed only from elementary exchange matrices and their inverses.

More complex braids of quasi-particles trajectories arise considering a quantum system consisting on more non-abelian anyons. An important note to mention here is that since quasi-particles are created and annihilated (fused) in pairs, only quantum systems of **even number of anyons** must be considered.

The exchange matrices notation in case of multi-anyons quantum systems has to account not only for the type of exchange of positions (clockwise or counterclockwise), but also it has to indicate the exact pair of anyons that swap positions.

To exemplify recall the braid of stocks prices in the figure 4 and leave aside the last strand in order to have an even number of strands that can be simulated by quasi-particles trajectories, such that the result is a 6 strands braid. Consider a quantum system of 6 quasi-particles, meaning 3 pairs of anyons, whose trajectories will simulate the strands of the braid.

The precise form of the braid in figure 4 can be executed by manipulating first the quasi-particles 4 and 5 in a counterclockwise swap of positions, which is defined by the matrix $\rho(\sigma_4)$. Next the anyons 1 and 3 are exchange their positions clockwise, a move that is expressed as the matrix $\rho(\sigma_1^{-1})$, which is the inverse of the matrix $\rho(\sigma_1)$. Finally quasi-particles 3 and 4 are forced to swap positions clockwise such that matrix $\rho(\sigma_3^{-1})$ reflects this last action taken over the 6-anyons quantum system.

Figure 6 shows the sequence of quasi-particles exchange positions maneuvers explained above, along with the braid that is formed by their trajectories.

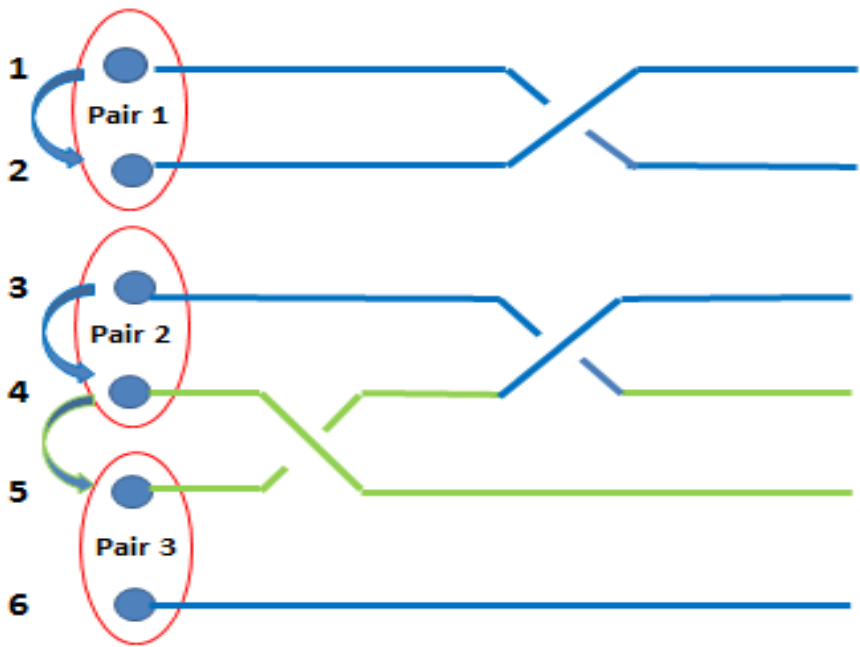

**Figure 6.** The braid that results manipulating 3 pairs of quasi-particles by exchanging their positions in a predefined sequence

The sequence of crossings in the braid executed by swapping the positions of quasi-particles and expressed as elementary exchange matrices can be written as:

$$\rho(\sigma_4)\rho(\sigma_1^{-1})\rho(\sigma_3^{-1}) \tag{3}$$

In topological quantum computation a sequence of braid matrices, like the above example implements quantum gates, the building blocks of quantum circuits and quantum algorithms. The matrix that results by multiplying the elementary exchange matrices and their inverses in a predefined order is the expression of a certain quantum gate, as it will be exemplified in the next sections.

An important particularity of the elementary exchange matrices is revealed by trying to answer the apparently trivial questions: Is the form of elementary braid matrices different as they appear by crossings of different adjacent strands on the braid of the quasi-particles trajectories? The answer is definitely YES and is a consequence of the fact that quasi-particles are created and fused only in pairs. Not getting deeper into physics aspects that partially will become clear in the next section, note here that the braid matrices resulting from swapping the positions of quasi-particles belonging to the same initial pair differ from the exchange matrices that result in exchanging the positions of anyons belonging to different initial pairs.

For the sake of exemplification, in figure 6 the braid matrix $\rho(\sigma_4)$ express the exchange position of quasi-particles belonging to different pairs since anyon 4 belong to initial pair 2 and anyon 5 to pair 3; the braided trajectories are colored in green to highlight this result. The trajectories of quasi-particles belonging to the same initial pair (pair 1 and respectively pair 2), swapping their positions defined by matrices $\rho(\sigma_1^{-1})$ and $\rho(\sigma_3^{-1})$ in figure 6, remained colored in blue. The three braid matrices have different form as it will be investigated in the following sections.

The explicit form of elementary exchange matrices also depends on the type of non-abelian anyons quantum computation is based on. From the two known topological quantum computer models, namely Fibonacci anyons and Ising anyons, the latter is chosen to simulate the braids of stocks and to realize quantum gates in the stock portfolio structure, since it has the braid matrices form more appropriate to the purpose of the further discussion. The Ising anyons model seems also to be more reliable since it is the most stable.

Next section provides information on how to construct qubits, the elementary quantum gates in the specification of quantum computing Ising anyons model and the effective form of the braid matrices that will be necessary to realize quantum gates on the stocks prices time series reunited in portfolios.

## 4. Ising anyons model realization of qubits and quantum gates

It should be emphasized since the beginning that this section has not the intention to be a complete guide to Ising anyons model of topological quantum computation. In the remaining sections, the present exposition closely follow the work of L. Georgiev [8],[9], [10], use sketches from Fan and co. [5] and remarks from M. Freedman and co. [7]. The interested reader is encouraged to look for in-depth details in the work of these remarkable physicists. Here only the remarks related exclusively to realization of quantum gates for stock portfolios are considered.

Quantum gates are typically represented as matrices. A quantum gate which acts on **k** qubits is represented by a **$2^k$ x $2^k$** unitary matrix. For the one-qubit quantum gates that will be analyzed here, the matrix is of the 2 x 2 form.

The most basic quantum system, **one-qubit basis system**, can be realized in the Ising anyons model by two pairs of quasi-particles. The two values of the qubit, |0> and |1> are constructed from the 4 quasi-particles that form the quantum system. Such one-qubit basis is exemplified in figure 7.

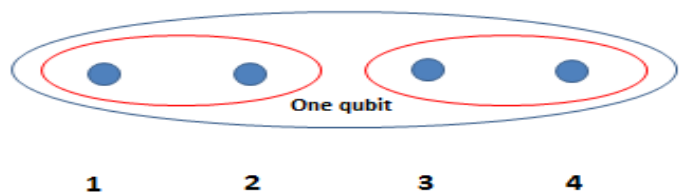

**Figure 7**. Two pairs of quasi-particles that can realize one-qubit basis quantum system

The single-qubit system sustains the realization of one-qubit quantum gates by swapping the positions of quasi-particles expressed as elementary exchange matrices. Since the braid matrices refer to one-qubit basis system their dimension must be 2 x 2. The explicit physical construction of braid matrices out of the exchange positions of the quasi-particles wave functions is leaved aside since is out of the scope of the present paper. It is suffice here to give only the final form of the braid matrices in the hypothesis of Ising anyons model.

Exchanging position of quasi-particles 1 and 2 (or 3 and 4), since they are coming from the same pair, as is shown in figure 7, is expressed by the braid matrix:

$$\rho(\sigma_1) = \rho(\sigma_3) = \begin{pmatrix} 1 & 0 \\ 0 & i \end{pmatrix} \quad (4)$$

Coming from different pairs, swapping the position of quasi-particles 2 and 3 results in the elementary exchange matrix:

$$\rho(\sigma_2) = \frac{e^{i\frac{\pi}{4}}}{\sqrt{2}} \begin{pmatrix} 1 & -i \\ -i & 1 \end{pmatrix} \quad (5)$$

All quantum gates are implemented by joining a series of these two elementary braid matrices and their inverses in a predefined sequence.

In quantum computation, one of the representative quantum gates is **Hadamard gate**. It will appear many times in the further exposition, such that it is the best example of how a quantum gate is actually constructed from a sequence of braid matrices.

Hadamard gate is represented by the **Hadamard matrix**:

$$H = \frac{1}{\sqrt{2}} \begin{pmatrix} 1 & 1 \\ 1 & -1 \end{pmatrix} \quad (6)$$

**Note here that in what follows the terms outside the matrix will be dropped, for the sake of simplicity, and only the form of the matrices will be retained.**

The series of elementary braid matrices needed to construct the Hadamard gate is:

$$\rho(\sigma_1)\rho(\sigma_2)\rho(\sigma_1) = \begin{pmatrix} 1 & 0 \\ 0 & i \end{pmatrix}\begin{pmatrix} 1 & -i \\ -i & 1 \end{pmatrix}\begin{pmatrix} 1 & 0 \\ 0 & i \end{pmatrix} \approx \begin{pmatrix} 1 & 1 \\ 1 & -1 \end{pmatrix} \quad (7)$$

and can be represented in the braid diagram that figure 8 shows:

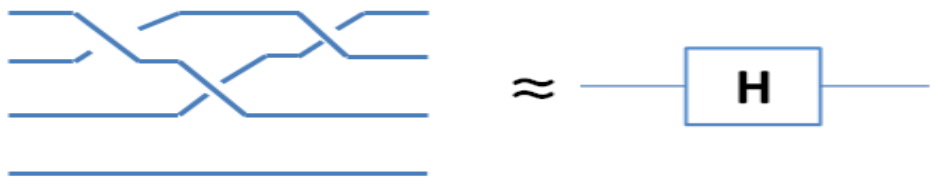

**Figure 8**. The Hadamard gate implemented as braiding trajectories of quasi-particles

where the symbol at the right is the consecrated representation of Hadamard gate in quantum computing literature.

Based on this suggestive example, it is easy to figure out, thinking of braid representation of stocks, how to identify quantum gates in the stock portfolio evolution.

### 5. Quantum computing simulation of stock portfolio

Prior to enter deeper in the problematic of finding quantum gates in stock portfolio evolution, some preliminary remarks on how to prepare the stocks for the simulation of prices time series by quasi-particle trajectories, are necessary.

In evaluation of stocks that will compose the portfolio subjected to realization of quantum gates, one important observation to mention is that prices of stocks which should be selected for simulation have to have close values in order to be relevant. This prescription for the price values of stocks in the portfolio assure the future high probability of seeing crossings between stocks prices time series that eventually lead to braid representations.

Recall the short fragment period of stocks prices time series chosen in the table 1; as of 3/19/2014 the prices of selected stocks varies in a restricted interval from the smallest 78,78 $ of **The Procter & Gamble Company (PG)** , to the bigger 96,1 $ of **McDonald's Corp. (MCD)**. The short interval for price values ensures the braid formation, as it is shown in table 2.

In contrast, choosing stocks having prices varying in a larger interval is irrelevant for quantum simulation since it is improbable to have crossings between their prices time series in a significant interval of time. To exemplify, consider the prices time series of stocks Cisco Systems Inc (CSCO) and Visa Inc (V) over the same period of time of 4 days from 3/19/2014 to 3/24/2014, as it is shown in the table 3.

| | | |
|---|---|---|
| 3/24/2014 | 21,57 | 220,75 |
| 3/21/2014 | 21,64 | 223,37 |
| 3/20/2014 | 21,83 | 221,82 |
| 3/19/2014 | 21,63 | 223,82 |
| | CSCO | V |

**Table 3.** Stocks having significantly different price values

Notice the big difference that exists between these two stocks prices making the occurrence of crossings between their prices time series almost impossible. A hypothetical portfolio containing these two stocks is absolutely irrelevant for quantum simulations.

A quantum computation simulation of a stock portfolio evolution starts by assigning a quasi-particle to each and every stock in the predefined portfolio. It is important to emphasize here that since the quasi-particles are created and fused in pairs, as it was stated in a latter section, the portfolio exposed to simulation must be composed by an **even number of stocks**. For the sake of exemplification it should be said that to simulate a stock portfolio evolution, from the fragment of 7 stocks selected in the section 2 only even number grater or equal to 4 have to be chosen.

The quasi-particles assigned to stocks are manipulated in exchanging positions clockwise and counterclockwise according to the braids occurring in the prices time series of stocks due to their quotations in the stock market. Once an overcrossing or an undercrossing appears between adjacent stocks the respectively assigned quasi-particles are swapping their positions counterclockwise or clockwise. Every swap of quasi-particles positions is reflected in a braid matrix. Predefined sequences of elementary braid matrices resulting in manipulating the anyons according to the stocks braids implements certain elementary quantum gates.

To exemplify the quantum gates realization in the stock portfolio evolution, fragmented periods are considered for the stocks prices time series, precisely the periods containing the braids necessary to realize a quantum gate. Cumulating many fragmented periods containing quantum gates in larger intervals of time form quantum circuits.

The realization of quantum gates will be directly exemplified for stocks prices braiding as they are sighted in the evolution of the stock portfolio.

## 6. One-qubit quantum gates of 4 stocks portfolio

The most basic quantum gates are constructed for single-qubit basis systems. In the Ising model of quantum computation single-qubit is expressed in terms of 4 quasi-particles. As it was pointed out in the latter section the braiding trajectories of the quasi-particles in the one-qubit basis representation are expressed as 2 x 2 elementary exchange matrices.

The elementary braid matrices for Ising anyons model in the one-qubit basis representation can be written as:

$$\rho(\sigma_1) = \rho(\sigma_3) = \begin{pmatrix} 1 & 0 \\ 0 & i \end{pmatrix}$$

$$\rho(\sigma_2) = \begin{pmatrix} 1 & -i \\ -i & 1 \end{pmatrix}$$

The construction of quantum gates out of these two elementary braid matrices is directly exemplified on the braid diagrams of portfolio consisting in 4 stocks. The 4 stocks are selected from *Dow Jones Industrial Average* stock components and the portfolio that will be analyzed is composed of: **The Procter & Gamble Company (PG)**, **The Walt Disney Company (DIS)**, **American Express Company (AXP)**, and **United Health Group Incorporated (UNH)**. The prices time series of these 4 stocks cover the 2014 whole year, but only fragmented periods are considered, periods that include a quantum gate realization.

The simplest elementary quantum gate that can be realized in the context of 4-stocks portfolio is the phase S-gate. The period selected for implementation of S-gate in the structure of 4-stocks portfolio is between 3/31/2014 and 4/1/2014, period that covers a braid over the stocks **DIS** and **PG** as it is shown in the table 4.

| | | | | |
|---|---|---|---|---|
| 4/1/2014 | 80.34 | 81.57 | 81.84 | 91.17 |
| 3/31/2014 | 80.07 | 80.6 | 81.99 | 90.03 |
| | dis | pg | unh | axp |

**Table 4**. A fragment period of stocks prices time series that exhibit a phase S-gate as the result of braided trajectories of stocks **DIS** and **PG**

The phase gate S is represented in quantum computation by the matrix:

$$S = \begin{pmatrix} 1 & 0 \\ 0 & i \end{pmatrix} \quad (8)$$

The braid matrix that Ising anyons model associate to braid generator $\sigma_1$ that can be noticed between stocks **DIS** and **PG** in the table 3, is

$$\rho(\sigma_1) = \begin{pmatrix} 1 & 0 \\ 0 & i \end{pmatrix}$$

which precisely simulates the quantum phase S-gate, having the same form as the S-gate matrix in relation (8).

Diagrammatically, the equivalence of the braid matrix $\rho(\sigma_1)$ with the phase S-gate:

$$\rho(\sigma_1) = \begin{pmatrix} 1 & 0 \\ 0 & i \end{pmatrix} = S \quad (9)$$

is represented as:

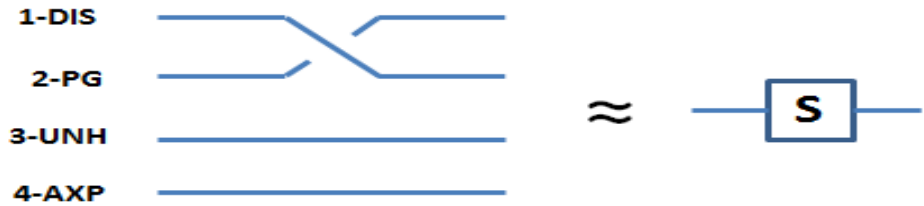

**Figure 9.** The phase S-gate is equivalent with braiding trajectories of **DIS** and **PG**

where in the right side is the consecrated symbol for the phase gate S.

The most important 1-qubit quantum gate is the **Hadamard gate.** The braid representation associated to this important quantum gate was formally introduced and discussed in section 4. The realization of Hadamard gate for the selected 4-stocks portfolio is exemplified by analyzing a period fragment from 3/18/2014 to 3/25/2014. This period covers the braid of stocks prices time series representing the Hadamard gate as can be depicted in the table 4.

|  | unh | pg | dis | axp |
|---|---|---|---|---|
| 3/25/2014 | 79.55 | 79.81 | 81.1 | 90.84 |
| 3/24/2014 | 79.3 | 79.49 | 81.28 | 91.01 |
| 3/21/2014 | 77.88 | 80.35 | 81.34 | 91.52 |
| 3/20/2014 | 78.32 | 80.81 | 81.53 | 91.69 |
| 3/19/2014 | 78.78 | 79.96 | 80.52 | 90.73 |
| 3/18/2014 | 78 | 79.77 | 81.99 | 91.59 |

**Table 4**. Fragment period of stocks prices time series that implements the Hadamard gate

The braid diagram associated to the winding trajectories of stocks prices time series in the table 4 pointing out the exact crossings of stocks that implement the Hadamard gate is represented in the figure 9.

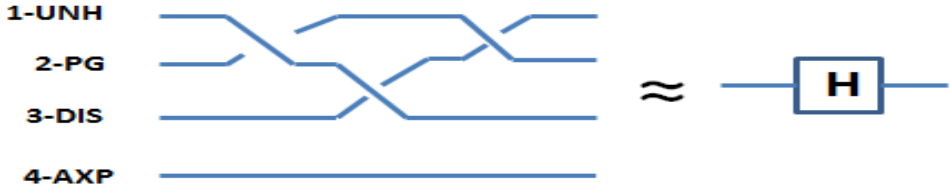

**Figure 9**. The Hadamard gate realization by braided trajectories of the 4-stocks portfolio

The series of elementary exchange matrices executing the Hadamard gate represented in the figure 9 is:

$$\rho(\sigma_1)\rho(\sigma_2)\rho(\sigma_1) = \begin{pmatrix} 1 & 0 \\ 0 & i \end{pmatrix}\begin{pmatrix} 1 & -i \\ -i & 1 \end{pmatrix}\begin{pmatrix} 1 & 0 \\ 0 & i \end{pmatrix} \approx \begin{pmatrix} 1 & 1 \\ 1 & -1 \end{pmatrix} \quad (10)$$

To specifically express the stocks that realize the Hadamard gate, the sequence of braid matrices above can be noted as:

$$\rho(\sigma_{UNH})\rho(\sigma_{PG})\rho(\sigma_{UNH}) \quad (11)$$

An **important observation should be mentioned here** is that the Hadamard gate can be executed by multiple sequences of braid matrices if the result of their multiplication is the same Hadamard matrix.

Consider in the table 5 below another fragment of stocks prices time series taken from 4/4/2014 to 4/16/2014.

| Date | pg | dis | unh | axp |
|---|---|---|---|---|
| 4/16/2014 | 78.19 | 78.95 | 81.65 | 87.4 |
| 4/15/2014 | 77.66 | 79.51 | 80.84 | 86.04 |
| 4/14/2014 | 77.62 | 79.18 | 80.81 | 85.5 |
| 4/11/2014 | 77.01 | 78.95 | 80.76 | 84.54 |
| 4/10/2014 | 77.51 | 79.99 | 81.09 | 85.36 |
| 4/9/2014 | 80.47 | 81.39 | 81.49 | 88.72 |
| 4/8/2014 | 79.57 | 80.66 | 81.35 | 86.49 |
| 4/7/2014 | 79.13 | 80.49 | 81.08 | 86.6 |
| 4/4/2014 | 79.77 | 80.43 | 81.53 | 89.17 |

**Table 5.** Another fragment period of stocks prices time series that implements the Hadamard gate

Figure 10 representing the braid diagram of stocks prices in table 5 is also realize a Hadamard gate.

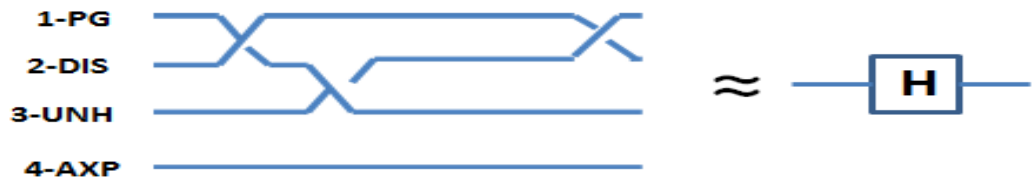

**Figure 10**. Realization of Hadamard gate by a different sequence of stocks prices braids

The accuracy of this statement can be easily verified by considering the sequence of elementary braid matrices that express the above braid diagram:

$$\rho(\sigma_1^{-1})\rho(\sigma_2)\rho(\sigma_1^{-1}) = \begin{pmatrix} i & 0 \\ 0 & 1 \end{pmatrix}\begin{pmatrix} 1 & -i \\ -i & 1 \end{pmatrix}\begin{pmatrix} i & 0 \\ 0 & 1 \end{pmatrix} \approx \begin{pmatrix} 1 & 1 \\ 1 & -1 \end{pmatrix} \quad (12)$$

which by neglecting the terms outside the matrices is precisely the Hadamard matrix. Another sequence of braid matrices that also implements the Hadamard gate is:

$$\rho(\sigma_1^{-1})\rho(\sigma_2^{-1})\rho(\sigma_1^{-1}) = \begin{pmatrix} i & 0 \\ 0 & 1 \end{pmatrix}\begin{pmatrix} 1 & i \\ i & 1 \end{pmatrix}\begin{pmatrix} i & 0 \\ 0 & 1 \end{pmatrix} \approx \begin{pmatrix} 1 & 1 \\ 1 & -1 \end{pmatrix} \quad (13)$$

An important category of quantum gates is the **Pauli gates**. The Pauli gates are three quantum gates namely X-gate, Y-gate and Z-gate, but only the first one will be exemplified here, leaving to the reader the pleasure of finding the other two.

To exemplify the realization of X-gate, consider the period between 5/9/2014 and 5/14/2014 as in the table 6.

| | | | | |
|---|---|---|---|---|
| 5/14/2014 | 77.17 | 80.92 | 81.17 | 88.46 |
| 5/13/2014 | 78.04 | 81.61 | 82.08 | 89.09 |
| 5/12/2014 | 77.74 | 81.73 | 82.42 | 89.66 |
| 5/9/2014  | 76.95 | 81.95 | 82.39 | 88.84 |
|           | unh   | dis   | pg    | axp   |

Table 6. The fragment period of stocks prices that leads to Pauli X-gate implementation

The X-gate is diagrammatically shown in the figure 11, although the winding of prices trajectories can already be hint from the table 6.

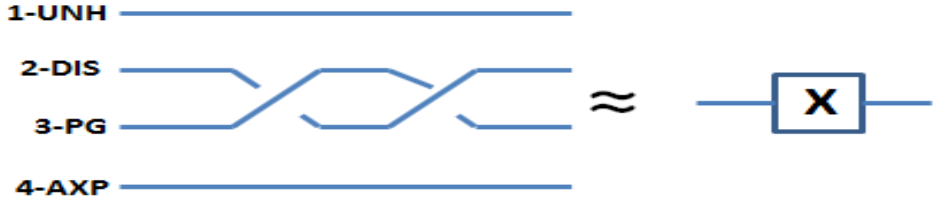

**Figure 11**. The representation of **Pauli X-gate** by the braided trajectories of stocks prices **DIS** and **PG**

The result can be verified by considering the sequence of elementary braid matrices from the figure above.

$$\rho(\sigma_2^{-1})\rho(\sigma_2^{-1}) = \begin{pmatrix} 1 & i \\ i & 1 \end{pmatrix}\begin{pmatrix} 1 & i \\ i & 1 \end{pmatrix} \approx \begin{pmatrix} 0 & 1 \\ 1 & 0 \end{pmatrix} = X \qquad (14)$$

The same X-gate can be realized considering the sequence

$$\rho(\sigma_2)\rho(\sigma_2) = \begin{pmatrix} 1 & -i \\ -i & 1 \end{pmatrix}\begin{pmatrix} 1 & -i \\ -i & 1 \end{pmatrix} \approx \begin{pmatrix} 0 & 1 \\ 1 & 0 \end{pmatrix} = X \qquad (15)$$

The Pauli X-gate is also known as the **quantum NOT gate**. Although the purpose of the present paper is to explore the realization of quantum gates for stock portfolios, the importance of the X-gate must be highlighted here. Previewing the work in the next paper it should be said that finding the X-gate in the portfolio evolution is equivalent to **shifting the trend of the stock portfolio** from up to down or from down to up.

The stock market has an incredible complexity; many quantum gates can be found in the winding of stocks prices trajectories for some periods of time. The tenacious reader can exercise to find some other one-qubit quantum gates for other portfolio of stocks, for other periods of time or for some other time frame quotations of prices, like minutes or even tick-by-tick prices.

What happen if consider the whole period from 3/18/2014 to 5/14/2014 covering all fragmented periods exemplified in this sections? The quantum gates individually found in the 4-stocks portfolio structure by analyzing fragments of time periods are matched together, the result being a **quantum circuit**, notion that will be investigated in the next section.

## 7. One-qubit quantum circuit of stock portfolio

Assembling together all the fragments of time periods covering the realization of one-qubit quantum gates discussed in the last section and taking a retrospective look over the whole period it results the table 7.

| Date | axp | unh | dis | pg | Gate |
|---|---|---|---|---|---|
| 5/14/2014 | 77.17 | 80.92 | 81.17 | 88.46 | X gate |
| 5/13/2014 | 78.04 | 81.61 | 82.08 | 89.09 | |
| 5/12/2014 | 77.74 | 81.73 | 82.42 | 89.66 | |
| 5/9/2014 | 76.95 | 81.95 | 82.39 | 88.84 | |
| 5/8/2014 | 76.89 | 81.6 | 82.16 | 88.62 | |
| 5/7/2014 | 77.91 | 80.29 | 82.09 | 87.98 | |
| 5/6/2014 | 75.26 | 81.03 | 81.13 | 86.2 | |
| 4/21/2014 | 74.95 | 79.11 | 81.56 | 86.67 | |
| 4/17/2014 | 75.78 | 79.99 | 81.76 | 86.22 | |
| 4/16/2014 | 78.19 | 78.95 | 81.65 | 87.4 | H gate |
| 4/15/2014 | 77.66 | 79.51 | 80.84 | 86.04 | |
| 4/14/2014 | 77.62 | 79.18 | 80.81 | 85.5 | |
| 4/11/2014 | 77.01 | 78.95 | 80.76 | 84.54 | |
| 4/10/2014 | 77.51 | 79.99 | 81.09 | 85.36 | |
| 4/9/2014 | 80.47 | 81.39 | 81.49 | 88.72 | |
| 4/8/2014 | 79.57 | 80.66 | 81.35 | 86.49 | |
| 4/7/2014 | 79.13 | 80.49 | 81.08 | 86.6 | |
| 4/4/2014 | 79.77 | 80.43 | 81.53 | 89.17 | |
| 4/3/2014 | 80.1 | 81.69 | 82.25 | 90.98 | |
| 4/2/2014 | 80.13 | 81.61 | 81.67 | 90.4 | |
| 4/1/2014 | 80.34 | 81.57 | 81.84 | 91.17 | S gate |
| 3/31/2014 | 80.07 | 80.6 | 81.99 | 90.03 | |
| 3/28/2014 | 78.99 | 79.76 | 81.62 | 90.46 | |
| 3/27/2014 | 78.48 | 79.65 | 81.02 | 89.99 | |
| 3/26/2014 | 78.62 | 79.5 | 81.54 | 89.66 | |
| 3/25/2014 | 79.55 | 79.81 | 81.1 | 90.84 | |
| 3/24/2014 | 79.3 | 79.49 | 81.28 | 91.01 | |
| 3/21/2014 | 77.88 | 80.35 | 81.34 | 91.52 | H gate |
| 3/20/2014 | 78.32 | 80.81 | 81.53 | 91.69 | |
| 3/19/2014 | 78.78 | 79.96 | 80.52 | 90.73 | |
| 3/18/2014 | 78 | 79.77 | 81.99 | 91.59 | |

**Table 7**. The whole period covering the realization of one-qubit quantum gates on the 4-stocks portfolio

The quantum gates realized in the analyzed period are colored in different colors, each color represent a certain quantum gate to make their location easier along the braided time series of stocks prices.

Notice that in the interval between 4/1/2014 and 4/3/2014 there is a crossing of stocks **UNH** and **DIS** expressed as the sequence of braid matrices:

$$\rho(\sigma_2^{-1})\rho(\sigma_2) = \begin{pmatrix} 1 & i \\ i & 1 \end{pmatrix}\begin{pmatrix} 1 & -i \\ -i & 1 \end{pmatrix} \approx \begin{pmatrix} 1 & 0 \\ 0 & 1 \end{pmatrix} \qquad (16)$$

that can be neglected since it represents the **Identity matrix**.

The resulting diagram of cumulating together the quantum gates for the larger time interval in the table 7 is the **quantum circuit** in the figure 12. The time flows from the left to the right.

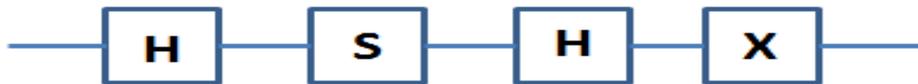

**Figure 12.** The quantum circuit that covers the period in the evolution of the 4-stocks portfolio

A quantum circuit is sequence of quantum gates. Typically a quantum circuit is designed to explicitly solve a certain complex problem. Quantum gates are arranged in a predefined sequence such that the expected output of the so formed quantum circuit to represent the answer to the input problem.

The issue when it comes to quantum circuit realized for the stock portfolios is that the quantum gates components are arbitrary implemented, being realized by price changes of stocks due to the stock market activity.

The sequence of quantum gates realized in the stock portfolio structure is not specifically destined to solve a particular complex problem. The output of a stock portfolio quantum circuit can only be guessed to refer at states of stocks, but the complete message remains unknown for now.

## 8. Two-qubit quantum gates of 6 stocks portfolio

Realization of two-qubit gates requires, in the Ising anyons model of quantum computation, a system of 6 quasi-particles. In the context of 6 quasi-particles system, the **first qubit** is constructed out of quasi-particles 1, 2, 3 and 4, and the **second qubit** from quasi-particles 3, 4, 5 and 6, as is depicted in the figure 13.

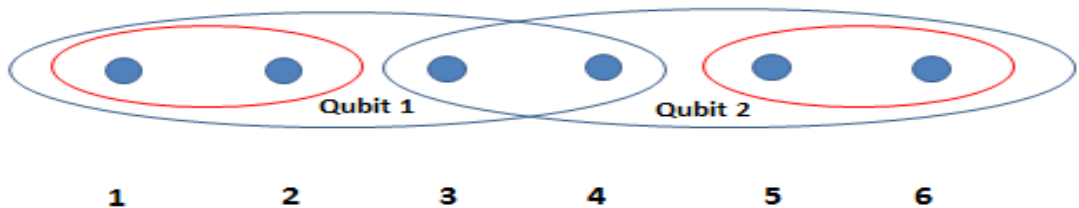

**Figure 13**. The 2-qubit basis system that can be realized with 6 quasi-particles in the Ising anyons model of quantum computing

In the figure 13 can be noticed that quasi-particles 3 and 4 are covering both the first qubit and the second qubit. These two quasi-particles entangle the two qubits of the system. The entanglement of the two qubits is the key in constructing the 2-qubits quantum gates. In the two-qubit basis system one of the qubit influences the other through the entanglement between them. This is the reason way the two-qubit quantum gates are known as being **controlled gates**.

Controlled quantum gates are realized in the Ising anyons model by braiding the trajectories of all the 6 quasi-particles that two-qubit system is composed of. In doing so all 6 anyons are exchange their positions in predefined sequences of exchange matrices capable to implement one specific 2-qubit quantum gate.

Another important reason to specifically identify the separation line between the two qubits representing the system since also one-qubit quantum gates can be embedded in two-qubit basis systems. The realization of embedded 1-qubit quantum gates needs only 4 quasi-particles from the total of 6 anyons of the system.

Returning to figure 13 the 1-qubit quantum gates in the two-qubit basis can be realized either by swapping the positions of the first 4 quasi-particles that reflects the first qubit or by exchanging the positions of the last 4 anyons that make up the second qubit. Simply speaking there are two different 1-qubit quantum gates that can be constructed in a 2-qubit basis system, one for the first qubit and one for the second qubit.

An odd behavior is signaled when trying to embed 1-qubit quantum gates in 2-qubit basis systems; the sequence of exchange matrices that implements a certain quantum gate differs from the first qubit to the second qubit. Say, the Hadamard gate realized on the first qubit have significantly different sequence of braid matrices than the Hadamard gate implemented on the second qubit.

It will be exemplified later in this section that the sequences of braid matrices which implement the same of 1-qubit quantum gate on first qubit and on second qubit are not equivalent.

In the 2-qubits basis system the elementary exchange matrices have $2^2$ x $2^2$ form, as it was specified in a latter section. Sequences of 4 x 4 elementary braid matrices implement both 1-qubit quantum gates and 2-qubits quantum gates.

Skipping again the details related to physics construction, the elementary braid matrices for the 6 quasi-particles can be written as:

$$\rho(\sigma_1) = \begin{pmatrix} 1 & 0 & 0 & 0 \\ 0 & 1 & 0 & 0 \\ 0 & 0 & i & 0 \\ 0 & 0 & 0 & i \end{pmatrix} \tag{17}$$

$$\rho(\sigma_2) = \begin{pmatrix} 1 & 0 & -i & 0 \\ 0 & 1 & 0 & -i \\ -i & 0 & 1 & 0 \\ 0 & -i & 0 & 1 \end{pmatrix} \tag{18}$$

$$\rho(\sigma_3) = \begin{pmatrix} 1 & 0 & 0 & 0 \\ 0 & i & 0 & 0 \\ 0 & 0 & i & 0 \\ 0 & 0 & 0 & 1 \end{pmatrix} \tag{19}$$

$$\rho(\sigma_4) = \begin{pmatrix} 1 & -i & 0 & 0 \\ -i & 1 & 0 & 0 \\ 0 & 0 & 1 & -i \\ 0 & 0 & -i & 1 \end{pmatrix} \tag{20}$$

$$\rho(\sigma_5) = \begin{pmatrix} 1 & 0 & 0 & 0 \\ 0 & i & 0 & 0 \\ 0 & 0 & 1 & 0 \\ 0 & 0 & 0 & i \end{pmatrix} \tag{21}$$

All two-qubit quantum gates are implemented from series of the above elementary braid matrices and their inverses.

The 2-qubits quantum gates are realized by braiding the trajectories of 6 quasi-particles, such that is natural to choose a **portfolio of 6 stocks** as basis in analyzing the quantum gates that stocks prices time series braids could reveal. The 6 stocks are again selected from *Dow Jones Industrial Average* components and the portfolio that will be simulated by the 6 quasi-particles is composed of: **The Walt Disney Company (DIS), Nike Inc. (NKE), McDonald's Corp. (MCD), American Express Company (AXP), UnitedHealth Group Incorporated (UNH), The Home Depot, Inc. (HD)**.

To exemplify realization of embedded 1-qubit quantum gates and controlled quantum gates on the 6 stocks portfolio a fraction covering the period between 10/16/2014 until 11/05/2014 is selected and reflected in the table 8. As usual, the historical quotations are daily closing prices taken from Yahoo finance and cover the 2014 year.

| Date | axp | dis | unh | nke | hd | mcd |
|---|---|---|---|---|---|---|
| 11/5/2014 | 91 | 92.42 | 94.64 | 94.68 | 95.78 | 95.8 |
| 11/4/2014 | 90.35 | 91.38 | 93.95 | 94.47 | 94.98 | 95.96 |
| 11/3/2014 | 90.85 | 91.71 | 93.45 | 93.61 | 94.86 | 96.09 |
| 10/31/2014 | 89.95 | 91.38 | 92.97 | 93.73 | 95.01 | 97.52 |
| 10/30/2014 | 88.74 | 90.22 | 93 | 93.38 | 93.88 | 97.52 |
| 10/29/2014 | 88.34 | 89.53 | 91.95 | 92.73 | 92.96 | 96.42 |
| 10/28/2014 | 88.01 | 89.93 | 92.45 | 92.6 | 92.64 | 96.59 |
| 10/27/2014 | 86.63 | 88.45 | 91.64 | 92.01 | 92.13 | 95.47 |
| 10/24/2014 | 86.4 | 88.61 | 90.9 | 91.64 | 91.67 | 94.99 |
| 10/23/2014 | 85.61 | 87.99 | 90.37 | 90.66 | 91.02 | 94.8 |
| 10/22/2014 | 83.96 | 87.1 | 89.77 | 89.86 | 90.94 | 93.34 |
| 10/21/2014 | 85.64 | 87.54 | 90.64 | 90.95 | 91.01 | 94.2 |
| 10/20/2014 | 84.01 | 85.52 | 88.54 | 88.9 | 91.59 | 91.85 |
| 10/17/2014 | 82.58 | 83.83 | 87.18 | 88.18 | 90.24 | 91.04 |
| 10/16/2014 | 80.24 | 81.74 | 85.39 | 87.04 | 88.88 | 89.91 |

**Table 8**. Fragment period covering the price time series of 6-stocks portfolio implementing 1-qubit and 2-qubit quantum gates

The portfolio will be fragmented this time in periods of time and unlike the 1-qubit system, also in braids of stocks prices according to the realization of a specified quantum gate.

To make the presentation of stocks quantum gates unitary and to closely analyzing the whole spectrum of problems the 2-qubit basis system is endowed with the aspects related embedding the single-qubit gates in 2-qubit basis and realization of controlled quantum gates will be treated separately.

### 8.1. Single-qubit quantum gates in 2-qubit basis

The two-qubit basis system allows two different choices for 1-qubit quantum gates to be constructed according to the qubit the braid matrices act on. It was stated in the last section that differences occur between sequences of braid matrices that implement the same quantum gate on the two separate qubits. Prior to exemplify on stocks prices time series the implementation of 1-qubit quantum gates in 2-qubit basis system an example proving this differences should be discussed.

In the case of Hadamard gate, depending on the qubit it acts on, the two quantum gates that can be realized are:

- **The Hadamard gate $H_1$ acting on the qubit 1**

$$H_1 \approx \rho(\sigma_1^{-1})\rho(\sigma_2^{-1})\rho(\sigma_1^{-1}) = \begin{pmatrix} 1 & 0 & 1 & 0 \\ 0 & 1 & 0 & 1 \\ 1 & 0 & -1 & 0 \\ 0 & 1 & 0 & -1 \end{pmatrix} \quad (22)$$

- **The Hadamard gate $H_2$ acting on the qubit 2**

$$H_2 \approx \rho(\sigma_4)\rho(\sigma_5)\rho(\sigma_4) = \begin{pmatrix} 1 & 1 & 0 & 0 \\ 1 & -1 & 0 & 0 \\ 0 & 0 & 1 & 1 \\ 0 & 0 & 1 & -1 \end{pmatrix} \quad (23)$$

Notice from the above sequence of braid matrices which implement the two Hadamard gate, that $H_1$ gate is realized by the first strands of the braid, reflected in the first qubit, and $H_2$ gate by the last strands that are expressed as the second qubit. Is evident here the differences between the two Hadamard matrices that represent the same Hadamard quantum gate realized over the two different qubits.

To exemplify on the 6 stocks portfolio, return to the table 8 and select a fraction covering the period between 10/17/2014 until 11/05/2014. The Hadamard gate is realized by the last 3 stocks, **NKE**, **HD** and **MCD** such that it implements the Hadamard gate $H_2$ on the second qubit, as it is shown in the figure 14, where the other crossings of stocks prices are leaved aside for now.

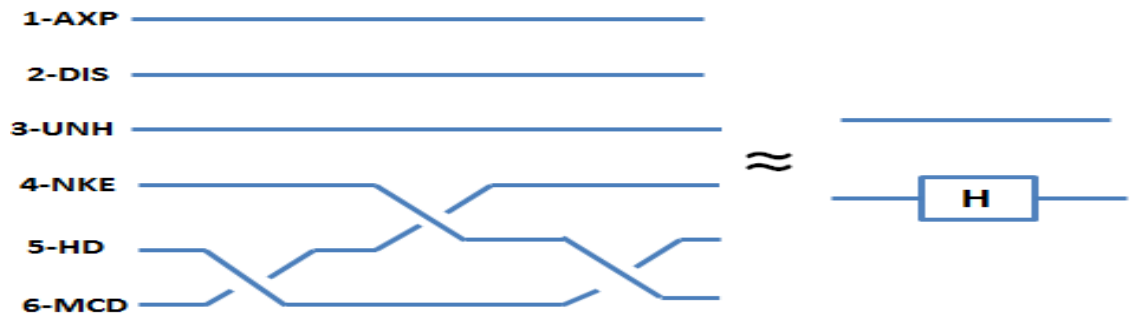

**Figure 14**. Realization of Hadamard gate $H_2$ on the second qubit of the system by braided prices time series of stocks **NKE**, **HD** and **MCD**

The sequence of elementary braid matrices leading to the realization of $H_2$ for the 6-stocks portfolio is:

$$\rho(\sigma_4)\rho(\sigma_5)\rho(\sigma_4) = \begin{pmatrix} 1 & -i & 0 & 0 \\ -i & 1 & 0 & 0 \\ 0 & 0 & 1 & -i \\ 0 & 0 & -i & 1 \end{pmatrix} \begin{pmatrix} 1 & 0 & 0 & 0 \\ 0 & i & 0 & 0 \\ 0 & 0 & 1 & 0 \\ 0 & 0 & 0 & i \end{pmatrix} \begin{pmatrix} 1 & -i & 0 & 0 \\ -i & 1 & 0 & 0 \\ 0 & 0 & 1 & -i \\ 0 & 0 & -i & 1 \end{pmatrix}$$

$$= \begin{pmatrix} 1 & 1 & 0 & 0 \\ 1 & -1 & 0 & 0 \\ 0 & 0 & 1 & 1 \\ 0 & 0 & 1 & -1 \end{pmatrix} \approx H_2 \qquad (24)$$

which is the Hadamard matrix $H_2$ acting on the second qubit in the context of 2-qubit basis system.

Returning to the table 8, take this time as reference the exchanges between stocks **UNH** and **NKE** in the interval between 10/16/2014 ad 10/20/2014. The series of braid matrices arising from the braids of the two stocks prices time series is equivalent with two parallel 1-qubit Pauli Z-gates executed on both first and second qubits:

$$\rho(\sigma_3)\rho(\sigma_3) = \begin{pmatrix} 1 & 0 & 0 & 0 \\ 0 & i & 0 & 0 \\ 0 & 0 & i & 0 \\ 0 & 0 & 0 & 1 \end{pmatrix} \begin{pmatrix} 1 & 0 & 0 & 0 \\ 0 & i & 0 & 0 \\ 0 & 0 & i & 0 \\ 0 & 0 & 0 & 1 \end{pmatrix} \approx Z_1 Z_2 \qquad (25)$$

The concept of parallel quantum gates is new in the presentation of stocks quantum gates, but for now it won't be investigate in deeper details.

The explicit braid diagram associated to the series of exchange matrices in the relation (25) is represented in the figure 15.

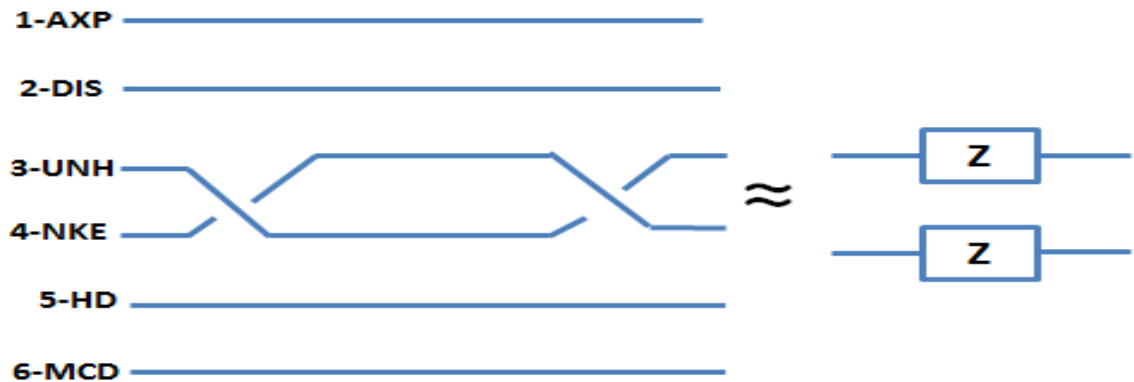

**Figure 15**. The parallel 1-qubit Z quantum gates that are realized by a sequence of two braids of stocks **UNH** and **NKE** prices

Finally the last quantum gate that can be realized from the braiding stocks in table 8 is the phase gate S. Exchanges between stocks **AXP** and **DIS** that took place between 11/3/2014 and 11/4/2014 are shown to implement the phase gate in figure 16.

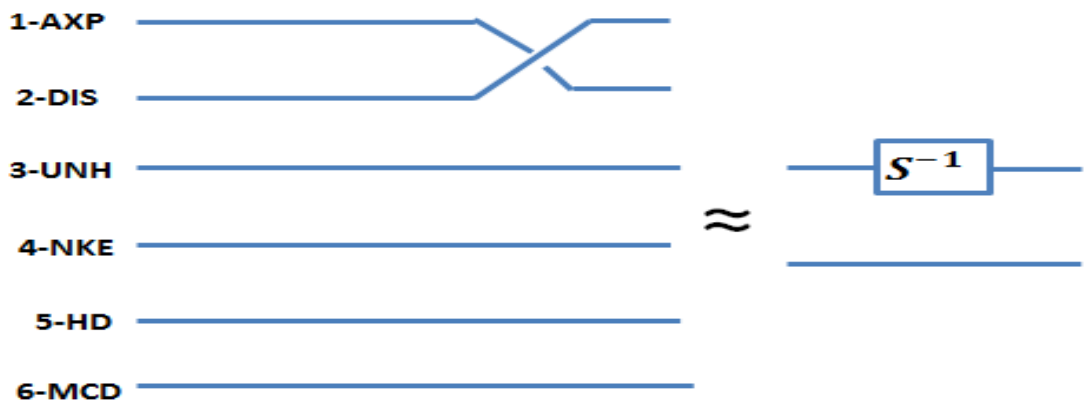

**Figure 16**. Implementation of an inverse S-gate for the first qubit in a two-qubit basis system represented by a 6-stocks portfolio

Notice that we have the inverse of $\rho(\sigma_1)$ braid matrix here, which represents the inverse of the phase gate S.

The 1-qubit quantum gates realized for a selected 6-stocks portfolio over a fragment period of time in this subsection are far from exhausting the possibilities of finding many other sequences of exchange matrices that implements quantum gates. Taking other periods of time for the same portfolio of stocks the other two Pauli matrices or some other configuration of quantum gates over the two-qubit basis might be found. For the purpose of exemplifying the existence of quantum gates in the structure of stock portfolios the examples analyzed here are suffice to form a comprehensive image.

### 8.2. Two-qubits controlled quantum gates

The two-qubit quantum gates act on both qubits of the system on the same time. In this two-qubit system one qubit is the controlled qubit while the other is the target qubit. In other words one qubit of the system influenced the other.

The most important two-qubit quantum gates are **controlled-Z (CZ)** and controlled-NOT (CNOT). From this two important quantum gates only the realization of controlled-Z gate will be exemplified on the 6 stocks portfolio.

The sequence of braid matrices implementing the CZ gate is:

$$\boldsymbol{CZ} = \rho(\sigma_1)\rho(\sigma_3^{-1})\rho(\sigma_5) = \begin{pmatrix} 1 & 0 & 0 & 0 \\ 0 & 1 & 0 & 0 \\ 0 & 0 & 1 & 0 \\ 0 & 0 & 0 & -1 \end{pmatrix} \qquad (26)$$

To find the precise series of braid matrices that realizes the CZ gate over the 6-stocks portfolio, a period between 11/05/2014 and 11/06/2015 is chosen and is shown in the table 9.

| | | | | | | |
|---|---|---|---|---|---|---|
| 11/6/2014 | 91.58 | 92 | 94.6 | 94.66 | 96.21 | 97.29 |
| 11/5/2014 | 91 | 92.42 | 94.64 | 94.68 | 95.78 | 95.8 |
| | dis | axp | mcd | nke | hd | unh |

**Table 9**. The implementation of controlled Z gate on the braids of all 6-stocks defining the portfolio

The observation that should be emphasizing here is that the same result may arise from other sequence of the above braid matrices since all three are diagonal. The interested reader is encouraged to search other sequences of braid matrices that implement a controlled Z gate in the structure of a stock portfolio.

The two-qubit controlled quantum gate Z symbol (at the right) along with the braid that it represents (at the left) is depicted in the figure 17.

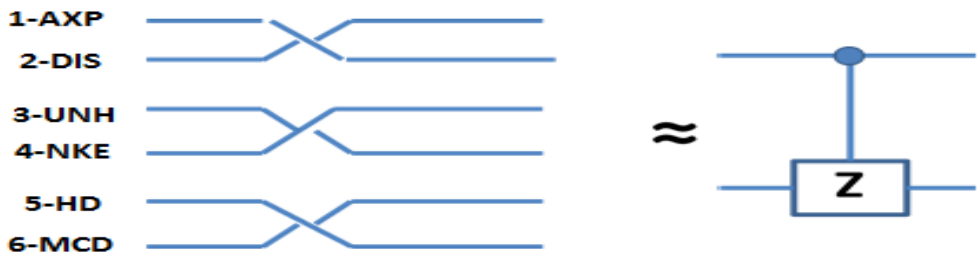

**Figure 17**. The controlled Z gate and the braid of stocks prices that it represents

As it was the case for single-qubit quantum gate, the series of embedded 1-qubit quantum gates and two-qubit quantum gates also form a quantum circuit that it will be investigated in the next section.

## 9. Two-qubit quantum circuits

The succession of 1-qubit and 2-qubit quantum gates implemented by the evolution of a 6-stocks portfolio can be assembled in a complex quantum circuit. For the period between 10/16/2014 and 11/5/2014 the price time series of the 6-stocks portfolio from the table 8 are braided to construct quantum gates that joined together realize the quantum circuit depicted in figure 18.

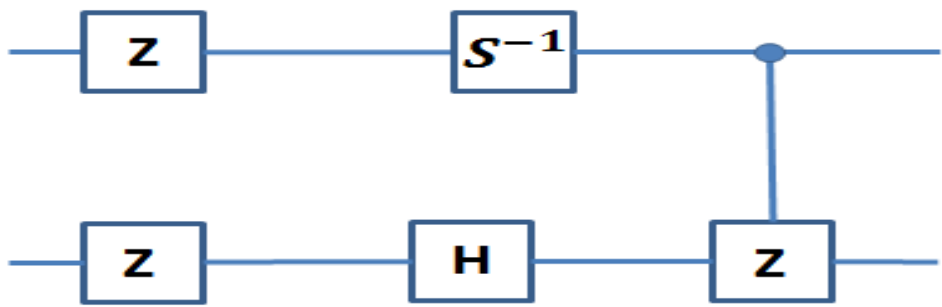

**Figure 18**. The quantum circuit constructed cumulating together the quantum gates realized in the hypothesis of the 2-qubit basis system

Denote here the increased complexity of the 2-qubit quantum circuit compared to that of the 1-qubit quantum circuit realized for the portfolio of 4-stocks. Still, the same observation that could as well applies here refers to

It might be helpful here to offer an alternative image of the quantum circuit above in order to make a parallel to today's realities in financial literature. Plotting the same time series of the 6-stocks of the portfolio in a chart that is sketched in the figure 19 the more familiar financial picture of cumulative time series is obtained.

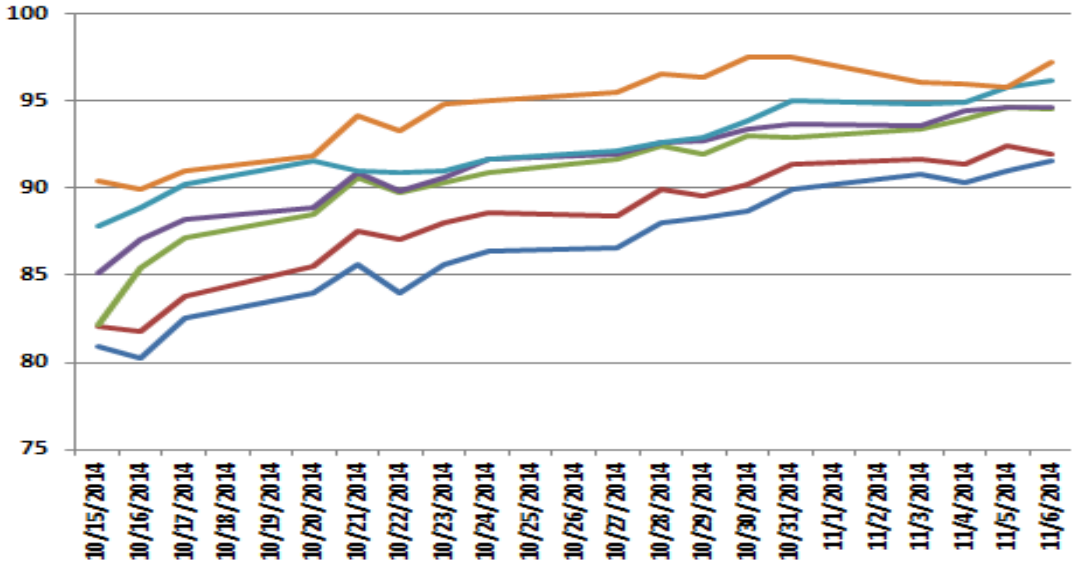

**Figure 19**. The chart of cumulative time series of stocks prices equivalent to the quantum circuit realized on the same period of time

Although it is difficult to put an equal sign between the quantum circuit and the cumulative time series chart, following the common sense in evaluating a stock portfolio, the last two figures refers to the same representation of the stock market.

## 10. n-qubit representation of stock market quantum code

Realization of quantum gates becomes more complex once new pairs of stocks are added to the initial portfolio. Difficulties in constructing the quantum gates occurs as new features arise by considering multi-qubit systems.

Not giving here specific examples from the stock market that will be investigated in a future paper a system of 3-quits basis is explored superficially. The Ising anyons model referring to 3-qubit basis is formed by 8 quasi-particles that braid their trajectories to realize quantum gates out of sequences of braid matrices.

In a 3-qubit basis, as it might be guess recalling the latter sections discussions, 1-qubit quantum gates can be embedded. Since the system consists of 3 qubits, three different 1-qubit quantum gates of the same kind could be realized.

Just for the sake of exemplification, the three one-qubit phase S gates that can be implemented in the 3-qubit basis system are directly expressed as single elementary 8 quasi-particles exchange matrices as:

$$S_1 = \rho(\sigma_1), \; S_2 = \rho(\sigma_5), \; S_3 = \rho(\sigma_7) \tag{27}$$

The elementary exchange matrices are not exemplified here since there are complicated $2^3 \times 2^3$ matrices that are out of the scope of the present paper.

The last of the three phase S-gates that can be implemented by 8 quasi-particles is represented in figure 20.

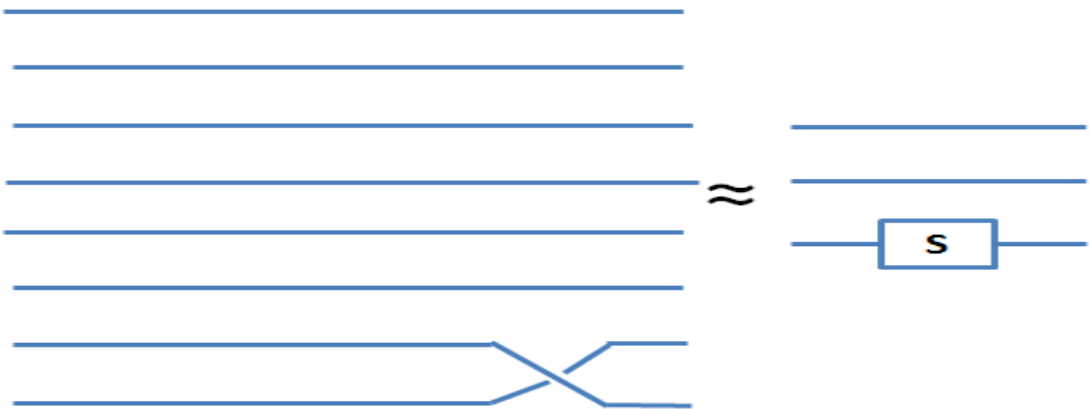

**Figure 20.** 1-qubit phase S-gate embedded in 3-qubit basis system

Having the practice of the last sections it is easy to find example of this simple quantum gate on the portfolio of eight stocks.

Difficulties appear when more complicated braids structures are considered. For example, the NOT gate $X_3$ acting on the third qubit has a different structure than the other two as it can be seen by writing the sequences of elementary exchange matrices:

$$X_1 = \rho(\sigma_2^2), \; X_2 = \rho(\sigma_4^2), \; X_3 = \rho(\sigma_4^2)\rho(\sigma_6^2) \tag{28}$$

Similarly, the Hadamard gates acting on the first and second qubits have different structures from their single-qubit counterpart H, while that acting on the third qubit cannot even be obtained exactly with the same number of elementary exchanges as H.

The element of novelty in the case of 3-qubit basis system is that embedding of 2-qubit quantum gates are also possible to realize.

Extending the scheme of finding quantum gates to **n-qubit systems** finally can result in simulating the entire stock market by expressing a market index like Dow Jones Industrial Average as an n-qubit quantum circuit. Although, extremely difficult to implement n-qubit quantum gates at the level of a market index, this endeavor is not impossible.

The complex quantum circuit resulting in joining the realized n-qubit quantum gates is the expression of a quantum code the stock market price moves is reflected on. This **mysterious quantum code of the stock market,** however, is arbitrary implemented due to the changings of stocks prices, and it not reflects, as in a typical quantum computation process, a certain problem to be solved. At this point it only can be argued that it might give indications about the future states of the stock market as the outcome of the realized quantum circuit.

In a future paper further steps to understand the **quantum code of the stock market** will be made.

## 11. Concluding remarks

The **beauty and magic of finance** that attract so many professionals over the stock exchanges around the world are exceeded only by the **greatness of quantum physics**. There are no other human fields of activity that would attract so much interest. When the two are combined a magical world is unfolded.

The "dance" of exotic subatomic particles, called non-abelian anyons, around each-other is netting braids that prove to have remarkable computation power. The topology of so created braids is reflected in quantum gates and quantum circuits through the sequences of elementary braid matrices.

Assimilated to quasi-particles trajectories, the trajectories of stocks prices are winding in braids that are also reflected in **braid matrices**.

To interpret the braiding of stocks the Ising anyons model of topological quantum computing is employed. This particular model of topological quantum computation was chosen for its relative simplicity of the braid matrices comparing with the other consecrated model, namely Fibonacci anyons model.

The Ising anyons model braid matrices are joined together to realize quantum gates in the prices time series of a stock portfolio structure. One by one the elementary quantum gates like the Hadamard gate, the Pauli gates and phase S gate are implemented for a stock portfolio composed of 4 stocks that simulates a one-qubit system.

Addition of another pair of stocks at the initial portfolios gives birth to a two-qubits system that can implement complex controlled quantum gates, like controlled Z gate and embedded 1-qubit quantum gates.

The scheme of finding quantum gates applies also to portfolios composed by many stocks that schematically are represented in n-qubit basis systems. Even a complex system like a market index composed by dozens of stocks can be represented by sequences of quantum gates acting on an n-qubit basis system.

Assembled together in the succession of their apparition on the stock portfolio evolution the quantum gates form quantum circuits. The complex quantum circuit that the stock components of a market index, such as Dow Jones Industrial Average, can realize in the succession of n-qubit quantum gates is reflected in a quantum code. Unlike the typical quantum computation where the sequence of quantum gates which make up the quantum algorithm is explicitly selected to solve a particular problem, the stock market quantum circuit is arbitrary. The quantum gates in the stock market arise as a result of stocks prices quotations changes.

Deciphering the **mysterious stock market quantum code** is an issue for future research.